# Scalable spin-photon entanglement by time-to-polarization conversion

Rui Vasconcelos[1,+], Sarah Reisenbauer[1,2,+], Cameron Salter[1,+], Georg Wachter[1,2], Daniel Wirtitsch[1,2], Jörg Schmiedmayer[2], Philip Walther[1], and Michael Trupke[1,2,*]

[1] *Faculty of Physics, University of Vienna, VCQ, Boltzmanngasse 5, 1090 Vienna, Austria*
[2] *Institute for Atomic and Subatomic Physics, Vienna University of Technology, VCQ, Stadionallee 2, 1020 Vienna, Austria*
[+] *These authors contributed equally to the work.*
[*] *Electronic address: michael.trupke@univie.ac.at*

**Spin-photon interfaces are strong candidates for building blocks of future quantum networks and quantum computers. Several systems currently under examination present promising features, but none of them yet fulfil all requirements for these aims[1–4]. A particularly attractive strategy for the realization of these applications is the creation of strings of entangled photons, where quantum correlations among the photons are mediated by operations on the spin of the emitter[5]. Here, we demonstrate for the first time the creation of spin-photon entanglement within the fundamental unit of a novel, scalable protocol based on time-to-polarization conversion. This principle allows us to bypass many of the imperfections of currently available photon sources and can therefore be utilized with a large variety of emitters. We execute the protocol using a nitrogen-vacancy centre in diamond, which possesses a long coherence lifetime and multiple spin degrees of freedom[6], thereby offering an outlook towards the creation of large entangled states[7].**

## INTRODUCTION

The generation of entangled photon states is of central importance in linear optical quantum computing (LOQC)[8,9] and optical quantum communication[10,11], and has potential applications in quantum sensing and metrology[12]. Currently, entangled photon sources rely mostly on spontaneous parametric down-conversion, which is robust and offers high purity, but is limited by intrinsically probabilistic entanglement generation[13]. For most applications in quantum technology, large entangled states are necessary in order to reach performance levels which exceed those of classical devices. The generation of such states therefore remains an outstanding challenge.

Cluster states are particularly desirable resources as they enable measurement-based quantum computation and have an in-built resilience to noise and loss[14–16]. Spin-based protocols have been developed for the generation of entangled photon strings, the most prominent of these being the "cluster state machine gun" (CSMG) of Lindner and Rudolph[5]. This protocol is appealingly simple and robust, and can be scaled to higher-dimensional cluster states using multiple spins[7,17]. Its requirements are however quite stringent. Ideally, it necessitates two orthogonally polarized, energy-degenerate optical transitions with negligible cross-decay terms (see Fig. 1a)). A first demonstration of the CSMG protocol with quantum dots was shown by Schwartz et al.[18], where the length of the cluster state was limited to three photons due to the short lifetime of the qubit.

Here we employ a nitrogen-vacancy (NV) defect in diamond, which is known for its excellent quantum coherence properties, and is therefore a promising candidate for quantum technology[6,19]. As a CSMG source, however, it presents several drawbacks. In particular, the optical transition energies connecting different spin states require careful tuning into resonance while maintaining negligible state mixing, and several of its excited-state levels can decay to multiple ground states and have a strong decay channel into a long-lived metastable state[20] (see Fig. 1b and 1c). It is therefore challenging to adhere to the requirements of the original CSMG scheme[21].

In this work we develop and demonstrate a novel, scalable scheme based on time-to-polarization conversion (TPC). Its main advantage is that it requires the use of just one optical transition (Fig. 1d). Therefore the most favourable spin properties of the strongest transition available can be used, and tuning of energy levels is no longer required. With these simplifications, it is applicable to a large variety of emitters.

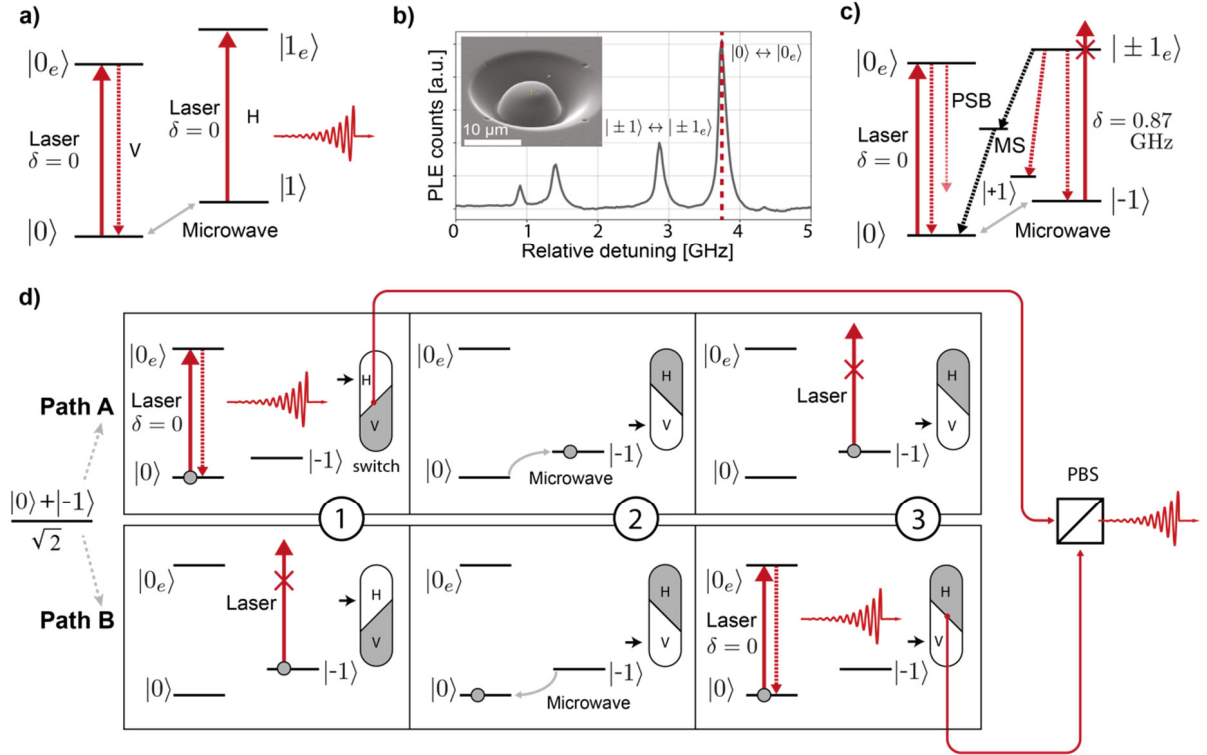

**Figure 1:** Conceptual overview. **a)** Ideal level scheme for the protocol of Lindner and Rudolph. Simultaneous excitation of two energy-degenerate transitions creates photons of orthogonal polarizations. Starting the cycle in an equal superposition of the states |0⟩ and |1⟩ creates a spin-photon entangled state of the type $(|0,V⟩ + e^{i\phi}|1,H⟩)/\sqrt{2}$ in the first iteration. **b)** Photoluminescence excitation spectrum of the NV centre used herein. Inset: electron microscope image of a solid-immersion lens milled into diamond. **c)** Simplified level scheme corresponding to the optical NV transitions. The emitter does not present the required energy-degenerate transitions with orthogonal photon polarization and undesired decay channels occur, particularly from the level $|\pm 1_e⟩$, which can non-radiatively decay to the metastable state (MS). Spin readout is performed by observing phonon sideband (PSB) photons. **d)** Ideal time-to-polarization entangling (TPC) scheme. The two ground states |0⟩ and |-1⟩ constitute the matter qubit. The red laser is resonant with the $|0⟩\leftrightarrow|0_e⟩$ transition and far-detuned (by 0.87 GHz in this experiment) from the $(|\pm 1⟩ \leftrightarrow |\pm 1_e⟩)$ transition, ensuring negligible cross excitation (see panel c). The two paths "A" and "B" display the evolution for initialization of the spin in the states |0⟩ and |−1⟩, respectively, as we run through our entangling sequence (from left to right). We start with the spin in the superposition $(|0⟩ + |−1⟩)/\sqrt{2}$ such that paths "A" and "B" also occur in superposition, forming the basis of the entanglement generation. In panels ①, a laser pulse is applied which leads to the generation of a photon in path "A". The photon is guided via a switch into a channel in which its polarization state is rotated to a horizontal orientation. No photon is generated in path "B" since no resonant transition is available from the state |−1⟩. In panels ②, the spin state is inverted in both paths by a microwave pulse and the classical optical routing is switched to a vertically polarized channel. In panels ③, a further excitation cycle is triggered with a second laser pulse. Due to the route switching, this excitation results in the generation of a vertically polarized photon in the evolution path "B". This time, no photon can be generated in path "A". Although the photon is originally emitted with the same polarization state in both evolution paths "A" and "B", the classical time-dependent photonic routing and polarization elements enable the entanglement of the spin with the photon polarization. The path information is erased with a polarizing beamsplitter at which the possible photon trajectories are overlapped simultaneously, resulting again in the desired entangled state of the type $(|0,V⟩ + e^{i\phi}|1,H⟩)/\sqrt{2}$ (see text). Together, the three-level system and the photonic routing elements therefore act as the desired four level system. The procedure can thus be iterated to generate a string of entangled photons, analogously to the CSMG protocol.

Briefly, our protocol starts with an equal spin superposition between the $m_s$=0 and $m_s$=-1 optical ground states. Upon excitation, the emitter produces a photon with 50% probability, which is stored in an *H*-polarized channel. During the storage time the superposition of the spin is inverted by a π-rotation. A second excitation pulse can then launch a photon, again with 50% probability, into a *V*-polarized channel. The sequence therefore generates a single photon in a superposition of polarizations. At this stage, it is however not possible to verify entanglement on the spin-photon state as the initial spin state is revealed directly by the photon's position.

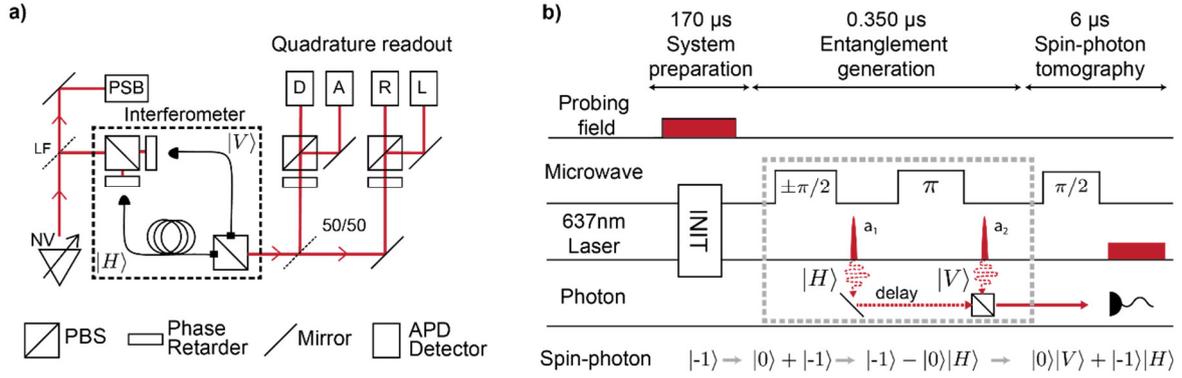

**Figure 2:** Overview of the experiment. **a)** Core elements of the photon collection path. A laser line filter (LF) separates the phonon-side band (PSB) photons from the zero-phonon line (ZPL) fluorescence path, where the spin-photon entanglement is detected. The TPC process is passively executed by the unbalanced interferometer, which consists of two polarization-maintaining single-mode fibers with a propagation time difference of 262 ns, mapping the wavepacket from the first (second) excitation/emission cycle into H (V) polarization and overlapping the two wavepackets at the output, where transmitted photons are monitored in quadrature, D/A and R/L. **b)** Experimental sequence to generate spin-photon entanglement by time-to-polarization qubit conversion (TPC). The dashed rectangle highlights the entangling operations, which can be iterated at will to scale the output to a multiphoton entangled state. The probing field is used to measure the phase of the interferometer. The microwave operations are rotations around the spin's y-axis. Events and states are shown on a time axis.

The path information, and with it the emission time information, is therefore erased by matching the delay time between excitation pulses to the storage time of the *H*-arm. After the TPC process, the time-separated wavepackets merge into a single polarization qubit entangled with the NV centre spin.

In more detail, the experimental sequence used to demonstrate the method unfolds in three steps: preparation, entanglement generation and tomography (Fig. 2):

**Preparation**: the experimental cycle (Fig. 2b) starts by initializing the NV charge state, through ionization into NV[-], with a green laser pulse. The electronic and nuclear ([14]N nucleus of the NV centre) spins are then initialized by iteratively flipping undesired spin populations using resonant optical pumping (see Supplementary information), preparing them in the $m_s = -1$, $m_I = 0$ state. The electron spin subspace $\{|0\rangle, |-1\rangle\}$ constitutes the matter qubit of the protocol. Initialization of the nuclear spin is applied to reduce dephasing. We obtain a qubit initialization fidelity of 97.9±1.6 % and a nuclear polarization, within the $m_s = -1$ manifold, of 83.8±1.9 %.

**Entanglement generation**: after initialization of the electron spin into $|-1\rangle$, a rotation with a microwave pulse $R_y(\pi/2)$ brings the electron into a superposition $\psi_0 = (|-1\rangle - |0\rangle)/\sqrt{2}$. An optical $|0\rangle \rightarrow |0_e\rangle$ π-pulse using a resonantly tuned laser then leads to the emission of a photon, conditional upon the state of the electron spin, resulting in the state $\psi(a_1) = (|-1\rangle|0_{a1}\rangle - |0\rangle|H_{a1}\rangle)/\sqrt{2}$. The ket $|H_{a1}\rangle$ ($|0_{a1}\rangle$) denotes an horizontally polarized photon (no photon) created by the pulse $a_1$. This photon is stored for 262 ns in the long arm of the fiber interferometer (Fig 2a). Meanwhile, a microwave π-pulse rotates the spin to the orthogonal state, $(|0\rangle|0_{a1}\rangle + |-1\rangle|H_{a1}\rangle)/\sqrt{2}$, followed by a second optical excitation ($a_2$), the emission of which is vertically polarized in the TPC apparatus, resulting in the entangled state $\psi(a_2) = (|0\rangle|V_{a2}\rangle + e^{i\phi}|-1\rangle|H_{a1}\rangle)/\sqrt{2}$, with the interferometer phase $\phi$ (see methods).

We can therefore realize a circuit analogous to the building block of the CSMG protocol by selecting the appropriate phase $\phi$ of the interferometer: Repetition of the entanglement generation step on the existing spin superposition leads to the addition of further photonic qubits to the entangled state (see Supplementary information). The correct routing of the photons is probabilistic here, and is heralded by the time of arrival, but can be made deterministic with an active switch.

**Tomography**: On the photonic side, the passive TPC scheme enables direct projection onto both the equatorial basis (D/A and R/L) in the path-erasing events, and onto the polar basis (H/V) by selecting path-revealing events (see Methods and Supplementary information). The D/A and R/L ports perform a projective measurement on the photon, dependent on the interferometer phase, onto the states:

$|\phi^\pm\rangle = |H\rangle \pm e^{i\phi}|V\rangle$ and $|\phi_{\pi/4}^\pm\rangle = |H\rangle \pm e^{i(\phi+\pi/4)}|V\rangle$. The electron spin readout relies on the rotation of the spin state through a $R_y(\theta)$ - pulse and a final 5 µs laser pulse (Fig. 2b).

We experimentally demonstrate the process using a NV centre in an artificial diamond created by chemical vapour deposition. A microlens (Fig. 1b) is machined over a pre-allocated NV centre by focused-ion beam milling for improved photon collection efficiency. We manipulate the spin using a microwave field radiated from two bond wires. The diamond is cooled to ~ 4.5 K in a closed-cycle cryostat and photons are collected through a window using a microscope objective.

## RESULTS

The fundamental unit of the proposed protocol is experimentally demonstrated by performing partial tomography on the resulting spin-photon system and quantifying the respective entanglement. The resulting measured correlations in the $\sigma_z \otimes \sigma_z$ basis are represented in Figure 3a, with a correlation value of $C_{zz} = \langle \sigma_z \otimes \sigma_z \rangle$ = (83.7±1.6)%.

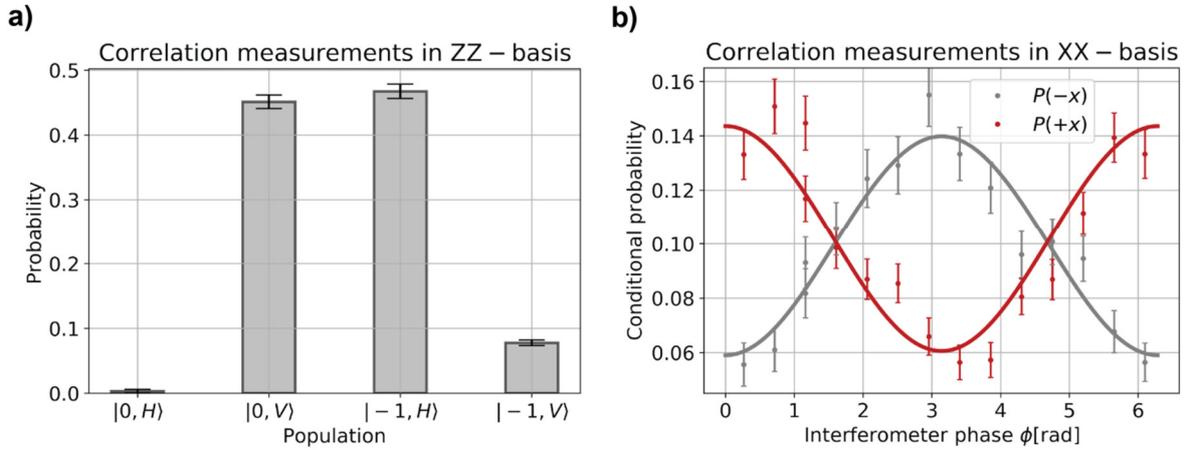

**Figure 3:** Characterisation of entanglement in orthogonal bases. **a)** Measured correlations in the $\sigma_Z \otimes \sigma_Z$ basis. In the passive TPC scheme, photons which are generated in the first (second) excitation cycle and take the short (long) path enable direct projection in the $|0/-1\rangle$ and $|H/V\rangle$ bases (see methods). These photons enable the reconstruction of the diagonal density matrix elements of the spin-photon system. **b)** Measured correlations in the $\sigma_X \otimes \sigma_X$ basis. Photons which arrive at the path-erasing time slot are expected to reveal a polarization correlation with the spin state in the equatorial bases. We plot the conditional probability of measuring the electron spin in the $|\pm X\rangle$ basis, upon detection of a ZPL photon, for a given interferometer phase $\phi$, which sets the photon measurement basis to $|\phi\rangle = (|H\rangle + e^{i\phi}|V\rangle)/\sqrt{2}$. The red (grey) points show the measurement data for preparation of the entangled states $|\psi\rangle = (|0\rangle|V\rangle + e^{i\phi}|-1\rangle|H\rangle)/\sqrt{2}$, with corresponding sinusoidal fits (lines). The clear deviation from a flat line witnesses the entanglement generation. The phase uncertainty for each data point is ±0.18 radians, as determined from the resolution of the phase readout (see Supplementary Information).

To extract the $C_{XX} = \langle \sigma_X \otimes \sigma_X \rangle$ correlations, spin-photon entanglement is generated for two different initial spin superposition states $\psi_- = (|-1\rangle - |0\rangle)/\sqrt{2}$ and $\psi_+ = (|-1\rangle + |0\rangle)\sqrt{2}$, corresponding to the spin measurement projection states $|\pm X\rangle$. Following established procedure[22], we now observe the spin-photon correlations by measuring their dependence on the interferometer phase, $\phi$. The photon detection events in the quadrature ports (D, A, R and L) are sorted according to $\phi$ at the time of detection and combined taking into account each port's phase offset. Fig. 3b shows the conditional probability of projecting the spin state onto $|+x\rangle$ and $|-x\rangle$, given the measurement of a photon in the equatorial basis $|\phi\rangle = |H\rangle + e^{i\phi}|V\rangle$. The resulting curves correspond to a correlation $C_{XX}$ =(40.7±2.9) %, showing the entanglement signature expected for the $|\psi^+\rangle$ Bell-state (as opposed to $C_{XX}$ =0 for a statistical mixture). In order to probe the quality of our source directly, the presented results have an accurately calibrated measure of the background light present in our ZPL detection window deducted (see Supplementary Information).

From the retrieved correlations, we estimate a lower bound on the entanglement fidelity, with respect to the ideal $|\psi^+\rangle$ Bell-state, of $F \geq$(64.7±1.3) % after (≥56.0±0.9) % before) background subtraction. This value is significantly above the bound for a classical state ($F \leq 50$ %), thereby demonstrating the entanglement in our spin-photon protocol, by over 11 standard deviations (over 6 without background

subtraction). The fidelity is currently limited by a variety of imperfections (such as spin mixing in the excited state manifold and imperfect spin readout) which can be minimized by improvements to the setup and system (see Supplementary information).

## DISCUSSION

The generation of entangled photon strings from quantum emitters will enable a multitude of applications in quantum technology. The TPC technique presented here relaxes many of the requirements placed on the emitter and will therefore broaden the range of systems for which such generation schemes are possible. We have performed the first demonstration of all key elements of the method, paving the path for future efforts towards generating multiphoton states. The emission and collection of photons from the optical transition can be drastically improved using an optical resonator. The most advanced results suggest that an enhancement of ZPL emission by a factor approaching 20 is possible with current technology[20]. Replacing the input PBS by a switch can further double the success probability per photon. Our initialization routine can be drastically shortened by including an additional laser on the $|\pm1\rangle \leftrightarrow |\pm1_e\rangle$ and implementing single-shot readout. With a feasible system efficiency of 40 % and a sequence duration of 10 μs, it will become possible to generate 3-photon cluster states at a rate exceeding 6 kHz, or 10-photon states at a rate of 10 Hz. NV centres are furthermore well-suited to the direct generation of two-dimensional cluster states, which are required for universal quantum computation[17]. The necessary quantum operations using ancillary spins[7] or remote centres have been impressively demonstrated in previous work[24,25].

## Materials and methods

### Sample and fabrication:

An artificial, single-crystal diamond of natural isotopic abundance and with a {1, 1, 1} surface orientation hosts the NV centre. We surveyed the diamond for shallow defects and created solid immersion lenses using focussed ion-beam milling over several defects with the desired N-V axis orientation (perpendicular to the surface). We then coated the surface with 110 nm of $SiO_2$ in order to reduce Fresnel reflection losses and laser backscatter at the high-index interface.

### Experimental details:

Our resonant optical pulses at around 637.2 nm are delivered from a narrowband external-cavity diode laser (Toptica DL Pro HP 637) and switched with two electro-optic amplitude modulators (Jenoptik AM635) in series. The fluorescence is split at a laser line filter into the resonant ZPL portion and the far off-resonant PSB portion. The latter was used to perform efficient spin readout. Since most of the photons decay into the PSB[26] and the system has significant photon losses, the measured ZPL efficiency (source to click) is ~2 × $10^{-5}$. The average probability of getting a spin readout click when prepared in $m_s = 0$ is 16.7±0.1%. We therefore observe an average of 36 PSB-ZPL coincidence events per hour.

The time-to-polarization conversion was performed by directing the ZPL part of the NV emission into a polarization-maintaining, fibre-based Mach-Zehnder interferometer. We matched the time between the two optical π-pulses in our sequence to the propagation delay between the arms of the interferometer. The interferometer was passively stabilized to minimize path length changes occurring during each entanglement cycle. The phase was furthermore tracked during each entanglement cycle by sending resonant laser pulses through the TPC between entanglement sequences and measuring the intensity on four quadrature detectors corresponding to the photon states *D*, *A*, *R* and *L*.

### Passive routing of photons:

The emitted photons are diagonally polarised with respect to the unbalanced interferometer´s arms and therefore split in two possible propagation paths (50% take the H(V)-polarised long (short) arm) and two corresponding propagation times. Therefore two excitation/emission cycles result in four arrival times at the output of the interferometer:

1) Early emission takes the short (V) path;
2) Early emission, long (H) path;
3) Late emission, short (V) path;
4) Late emission, long (H) path.

If the separation between the two emissions matches the time difference in the propagation, then events of type 2) and 3) have the same arrival time. Detection within this time window erases the path information and heralds the intended function of the passive switch (50% success probability). Events outside this window, 1) and 4), are path-revealing and are not conducive to the creation of the photonic polarisation qubit.

Electron and nuclear spin initialization:

The initialization sequence relies on electron and nuclear spin flips in the optically excited state of the NV centre. A low-power laser pulse (5 μs long), resonant with the $|0\rangle \leftrightarrow |0_e\rangle$ transition, results in an electron (nuclear) spin flip with high (low) probability. Subsequently, nuclear spin selective microwave pulses are applied to the ground state, which drive the population in the undesired states back to $|0\rangle$. This sequence is repeated several times in order to enhance the probability of initializing the electron (nuclear) spin in state $|m_s = -1\rangle$ ($|m_I = 0\rangle$).

Fidelity estimation:

Following a well-established method[22,25] we calculate the lower bound on the fidelity as:

$$F \geq 0.5 \left(\rho_{22} + \rho_{33} - 2\sqrt{\rho_{11}\rho_{44}} + C_{XX}\right)$$

where $\rho_{ii}$ denotes the diagonal entries of $\rho$ and $C_{XX}$ denotes the XX basis correlations $C_{XX} = \langle \sigma_X \otimes \sigma_X \rangle$ of $\rho$. The diagonal elements $\rho_{11}, \cdots, \rho_{44}$ are directly extracted from the data in Fig 3a, whereas the equatorial correlations $C_{XX}$ are calculated from the contrast of the curves in Fig 3b.

Further details of all methods are provided in the supplementary information.

Note: During preparation of the manuscript, we became aware of preliminary efforts towards the results achieved in this work using a quantum dot[27].

## Acknowledgements
We are grateful to the WWTF (project ICT12-041 PhoCluDi) the FWF (projects M 1852-N36 Lise Meitner Programme, W1210-N25 DK-COQUS, I 3167-N27 SiC-EiC, and W1210 DK-Solids4Fun), the TU Innovative Projekte, and the EU Marie Curie Actions project 628802. The focussed ion beam milling for this project was carried out using facilities at the University Service Centre for Transmission Electron Microscopy, Vienna University of Technology, Austria.

## Conflict of interests
The authors declare no competing interests.

## Contributions
R. V., S. R. and C. S. performed the experiments. R. V., S. R., C. S. and M. T. analyzed the data. G. W. and D. W. contributed to the experimental apparatus. J. S., P. W. and M. T. provided support for the work. M. T. devised the scheme, supervised the work and drafted the manuscript. All authors contributed to the interpretation of the data and the writing of the manuscript.